\documentstyle[aaspp4]{article}
\newcommand{\figcomment}[1]{#1}
\figcomment{
\input epsf
\def\jepsfbox#1{\typeout{#1} \epsfbox{#1}}
\def\plottwo#1#2{\centering \leavevmode
\epsfxsize=.45\columnwidth \jepsfbox{#1} \hfil
\epsfxsize=.45\columnwidth \jepsfbox{#2}}
}
\def\unsetyr{\def\oyear{\relax}\def\cyear{\relax}\def\cyeara{a\relax}\def\cyearb{b\relax}\def\cyearc{c\relax}\def\cyeard{d\relax}\def\cyeare{e\relax}\def\cyearf{f\relax}\def\cyearg{g\relax}}
\def\setyr{\def\oyear{(}\def\cyear{)}\def\cyeara{a)}\def\cyearb{b)}\def\cyearc{c)}\def\cyeard{d)}\def\cyeare{e)}\def\cyearf{f)}\def\cyearg{g)}}
\unsetyr
\def\jcite#1{\setyr\cite{#1}\unsetyr}
\def\rmmat#1{{\hbox{\rm #1}}}
\def\rmscr#1{\rmmat{\scriptsize #1}}
\newcommand{\be}{\begin{equation}}
\newcommand{\ee}{\end{equation}}
\newcommand{\ba}{\begin{eqnarray}}
\newcommand{\ea}{\end{eqnarray}}
\newcommand{\ie}{{\it i.e.~}}
\newcommand{\eg}{{\it e.g.~}}
\def\d{{\rm d}}
\def\dd#1#2{\frac{\d #1}{\d #2}}
\def\eqref#1{Equation~\ref{eq:#1}}
\def\figref#1{Figure~\ref{fig:#1}}
\begin{document}
\title{How Common Are Magnetars?  The Consequences of Magnetic-Field Decay}
\author{Jeremy S. Heyl}
\authoremail{jsheyl@tapir.caltech.edu}
\affil{Theoretical Astrophysics, mail code 130-33,
California Institute of Technology, Pasadena, CA 91125}
\author{S. R. Kulkarni}
\authoremail{srk@astro.caltech.edu}
\affil{Division of Physics, Mathematics and Astronomy, 105-24,
California Institute of Technology, Pasadena, CA 91125}

\begin{abstract}
Ultramagnetized neutron stars or magnetars have been invoked to
explain several astrophysical phenomena.  We examine how the magnetic
field of a magnetar will decay over time and how this decay affects
the cooling of the object.  We find that for sufficiently strong
nascent fields, field decay alters the cooling evolution significantly
relative to similarly magnetized neutron stars with constant fields.
As a result,  old magnetars can be expected to be bright in the
soft X-ray band. The soft X-ray source
RXJ~0720.4$-$3125 may well be the nearest such old magnetar.
\end{abstract}
\leftline{  stars: neutron --- stars: magnetic fields --- gamma rays:
observations }
\section{Introduction}

The birth parameters -- the rotation period ($P$), the strength of the
magnetic field ($B$) and the velocity -- of neutron stars are a matter
of fundamental interest in our study of neutron stars. Another equally
fundamental issue is the subsequent evolution of the magnetic
field.  About a decade ago, it was thought that most neutron stars are
born with parameters similar to that of Crab, initial period of tens of
milliseconds (perhaps to hundreds) and magnetic field strength confined
to a narrow range of $10^{11-12}$ G (\eg\ \cite{Nara87}).

Perhaps the single most important parameter is the strength of the
magnetic field, $B$. This strength determines the luminosity, the
lifetime and even the nature of the energy loss from isolated neutron
stars.  All these three issues affect observability and hence can be
ignored at the risk of severe selection. For example, \jcite{Usov96}
argue that when $B$ exceeds $0.1B_{cr}$ the pulsar wind may be
dominated by bound pairs rather than freely streaming electrons and
positrons; here $B_{cr}=m^2c^3/e\hbar \approx 4.4 \times 10^{13}$~G.
Such strongly magnetized neutron stars may thus be invisible in the radio
sky.

Over the past decade there has been increasing recognition that there
is considerable diversity in the magnetic field strengths of neutron
stars. In particular, there is circumstantial evidence suggesting the
existence of ``magnetars'' -- neutron stars with magnetic field
strengths in excess of $B_{cr}$ (\cite{Thom95}). The evidence
primarily comes from observations of soft gamma-ray repeaters (SGRs)
and long-period pulsars in supernova remnants, especially young
supernova remnants (\cite{Vasi97b}). \jcite{Kulk93}
have argued that SGRs are young neutron stars, albeit with
some special parameters, based on their associations with supernova
remnants. They and \jcite{Kouv94} estimate the birth rate of
SGRs is 10\% that of ordinary pulsars.
\def\rxj{RX~J$0720.4-3125$}

In this {\it Letter} we discuss the possibility of detection of nearby
old magnetars. Such soft X-ray emission is expected on various
theoretical grounds: \jcite{Thom96} through magnetic-field decay and
\jcite{Heyl97magnetar} by neutron-star cooling.  Here, we will examine
the consequences of magnetic-field decay in magnetars in the context
of detailed models of magnetized neutron-star envelopes
(\cite{Heyl98numens}).  The essential idea is as follows: the primary
store of energy in a magnetar is the energy stored in the magnetic
field, $E_B \sim B^2R^3$; here $R$ is the radius of the neutron star.
Assuming that the field decays on a time scale, $\tau_D$ we derive a
luminosity $E_B/\tau_D$. Some portion of this could heat up the
surface leading to potentially detectable soft X--ray emission.

The organization of the paper is as follows. In section~\ref{sec:rxj}
we discuss a candidate magnetar, \rxj\ (\cite{Habe97}). The importance
of this object is that it is nearby, a few hundred parsecs, and thus
must be derived from a substantial population (\cite{Kulk98};
hereafter KV). This is a crucial point in that it provides primary
observational motivation to this paper. In section~\ref{sec:decay} we
discuss potential field decay mechanisms and in section~\ref{sec:calc}
we calculate the expected thermal emission in old magnetars. In
contrast to earlier work (\eg \cite{Thom96}), these calculations take
account explicitly the modification of thermal conductivity by strong
magnetic fields ($B \geq 10^{11}$~G) (\cite{Hern84a,Heyl98numens}) and
its effect on the structure of the neutron-star envelope
(\cite{Heyl97magnetar}, \cite{Heyl98numens}).  Finally in
section~\ref{sec:discuss} we present estimates for the number of
nearby magnetars that are detectable solely based on their thermal
emission powered by magnetic field decay.

\section{Soft X-ray emission from Magnetar Candidates}
\label{sec:rxj}

\jcite{Thom96} argue that the soft X-ray counterpart of SGR 0526$-$66
(\cite{Roth94}) is powered by magnetic field decay.  It is in this
connection that the nearby soft X-ray source, \rxj\ has drawn
considerable attention. The source -- a bright, soft X-ray ($kT=80$
eV) pulsar, $P=8.39$ s -- was identified in the ROSAT All Sky-Survey
by \jcite{Habe97}. \jcite{Motc98} and KV searched for an optical
counterpart for this source.  KV identify a plausible faint blue
counterpart. In any case, Motch \&\ Haberl and KV argue that the
optical counterpart is so faint that \rxj\ is most likely a nearby
(100 pc) isolated neutron star.

KV and \jcite{Heyl98rxj} have discussed in detail the possibility that
\rxj\ is a nearby magnetar.  Its soft X-ray emission may be powered by
the decay of its magnetic field.

\section{Avenues for Magnetic-Field Decay}
\label{sec:decay}

\jcite{Gold92} studied several avenues for magnetic field decay in
isolated neutron stars: Ohmic decay, ambipolar diffusion and Hall
drift.  Depending on the strength of the magnetic field, each of these
processes may dominate the evolution:
\def\rhorhon{\left ( \frac{\rho}{\rho_\rmscr{nuc}} \right )}
\def\yr{\rmmat{~yr}}
\ba
t_\rmscr{Ohmic} & \sim & 2 \times 10^{11} \frac{L_5^2}{T_8^2}
\rhorhon^3 \yr \\
t^\rmscr{s}_\rmscr{ambip} & \sim & 3 \times 10^{9} \frac{L_5^2
T_8^2}{B_{12}^2} \yr \\
t^\rmscr{irr}_\rmscr{ambip} & \sim & \frac{5 \times 10^{15}}{T_8^6
B_{12}^2} \yr + t^\rmscr{s}_\rmscr{ambip} \\
t_\rmscr{Hall} & \sim & 5 \times 10^{8} \frac{L_5^2
T_8^2}{B_{12}} \rhorhon \yr 
\ea
where $L_5$ is a characteristic length scale of the flux loops through
the outer core in units of $10^5$~cm, $T_8$ is the core temperature in
units of $10^8$~K and $B_{12}$ is the field strength in units of
$10^{12}$~G.  Ohmic decay dominates in weakly magnetized neutron stars
($B\lesssim 10^{11}$~G), fields of intermediate strength decay ($B
\sim 10^{12} - 10^{13}$~G) via Hall drift, and intense fields ($B
\gtrsim 10^{14}$~G) are mostly strongly affected by ambipolar
diffusion.  \jcite{Thom96} examine the possibility of obtaining an
equilibrium between neutrino cooling and heating through magnetic
field decay for $B \sim 10^{15}$~G and $T \sim 10^8$~K.

Ambipolar diffusion may be decomposed into a solenoidal component
which does not upset the equilibrium among the neutrons, protons and
electrons and an irrotational component.  The field evolution through
the irrotational component is accompanied by modified (or direct if
allowed) URCA reactions to restore the equilibrium; therefore, it
generally operates more slowly than the solenoidal component.  It is
difficult to construct topologically a solution for the ambipolar
diffusion of a decaying field without both components;
therefore, the rate of ambipolar diffusion is set by the slower of
the two processes.  A strong magnetic field may speed the irrotational
component by catalyzing the direct URCA process in the
out-of-equilibrium component (\cite{Lein97}); therefore, we treat both
the case where the fast solenoidal mode and where the slower
irrotational mode set the field decay timescale.

\section{Calculations}
\label{sec:calc}

To estimate the total energy released in the crust by the decay of the
magnetic field, we assume that the field is a dipole outside of the
neutron star and in the insulating envelope, and that the neutron star
crust is thin relative to the radius of the star.  The total energy
contained in the neutron star's magnetic field is estimated by
\be
E_B = \frac{1}{12} B_p^2 R^3
\ee
where $B_p$ is the strength of the magnetic field at the pole at the
surface, and $R$ is the radius of the neutron star.  Substantial
magnetic energy may be contained in small structures below the
envelope; consequently, this is a lower bound on the total energy of
the neutron star's magnetic field.  The power released by the decay of
the field is
\be
{\dot E}_B= -\frac{1}{6} \dd{B_p}{t} B_p R^3.
\ee
We will assume that the magnetic energy is dissipated below that
region of the crust which supplies the bulk of the opacity (\ie the
sensitivity region); the field decay heats the isothermal core of the
neutron star.

To calculate the thermal evolution of the neutron star, we use the
$T_\rmscr{c}-T_\rmscr{eff}$ relationship presented in \jcite{Heyl97magnetar}
using the conductivities derived by \jcite{Hern84a}, and extended to
weaker fields using the numerical techniques outlined by
\jcite{Heyl98numens}.  The envelope calculations do not include a
treatment of the Coulomb pressure terms in the equation of state (\eg
\cite{VanR88}).  Although this may introduce uncertainties to the
envelope calculations for low effective temperatures ($T_\rmscr{eff} <
10^6$~K), the correction to the $T_\rmscr{c}-T_\rmscr{eff}$ relationship due
to Coulomb interactions is probably not a large as implied by the
results of \jcite{VanR88}.  The modified Coulomb prescription used by
\jcite{VanR88} yield a large correction to the $T_\rmscr{c}-T_\rmscr{eff}$
relation for precisely those models which contain a region of
negative total pressure.

We use the same cooling model as \jcite{Heyl97magnetar} coupled with
an equation for the field decay of the form
\be
\dd{B_p}{t} = -B_p
\left ( \frac{1}{t_\rmscr{Ohmic}} + \frac{1}{t_\rmscr{ambip}} +
\frac{1}{t_\rmscr{Hall}} \right )
\ee
The middle term dominates the sum except during the first $\sim
500$~yr of the decay of the $10^{14}$~G field when Hall drift has a
comparably strong effect.  We study three initial field strengths
$B_{p,0}=10^{14}, 10^{15}$ and $10^{16}$~G, and begin the integrations
at $t=1$~yr and a core temperature of $10^9$~K until $t=10^7$~yr using
logarithmically spaced timesteps.

\figref{decay} shows the evolution of the effective temperature and
magnetic field strength for the various initial field strengths and
assumptions regarding the disposition of the magnetic energy.  The
release of magnetic energy heats the core, increasing the emergent
flux from the neutron star.

For the solenoidal mode, the effect is substantial for an initial
field of $10^{16}$~G at $t=10^3$~yr, the effective temperature is 40\%
higher than if the field did not decay.  For $B_0=10^{15}$~G, the
effective temperature is 10\% higher at $t=10^4$~yr.  Because the
initial magnetic field strength has not been assumed to be infinite,
the evolution of the magnetic field strength departs from a power law.
However, for the strongest initial field ($B_0=10^{16}$~G), the
evolution approaches a power law after a few characteristic decay
timescales.  The slope of the power law changes appreciably when the
neutron star leaves the neutrino-dominated cooling era, and photon
emission begins to dominate the cooling (at $t\sim 10^6$~yr).
Additionally, as the energy released in the field decay heats the
core, the decay is slowed slightly as apparent in the upper-right
panel of \figref{decay}.

The character of field decay in the irrotational mode is different.
Since the field tends to decay more slowly, its effect on the cooling
of the neutron star is most pronounced later in the star's evolution
($t \sim 10^6$~yr).  For $B_0=10^{16}$~G, the neutron star with a
decaying field is brighter throughout its evolution than a similar
neutron star with a static field.  For a more weakly magnetized
neutron star ($B_0=10^{15}$~G), we find that heating by the decaying
field and neutrino cooling attain a near equilibrium for $t\sim
10^5$~yr, similar to the result found by \jcite{Thom96}.  Regardless
of their initial fields as long as $B_0 \gtrsim 10^{15}$~G, neutron
stars with decaying fields are much brighter during the photon-cooling
epoch than their coevals with static fields.

For weaker fields however ($B\lesssim 10^{14}$~G), both the total
energy contained in the magnetic field and the rate of its decay are
insufficient to perturb the cooling evolution during this epoch.  Even
if ambipolar diffusion is unimportant, the field decay by Hall drift
can result in a significant increase in the surface emission.
\figcomment{
\begin{figure}
\plottwo{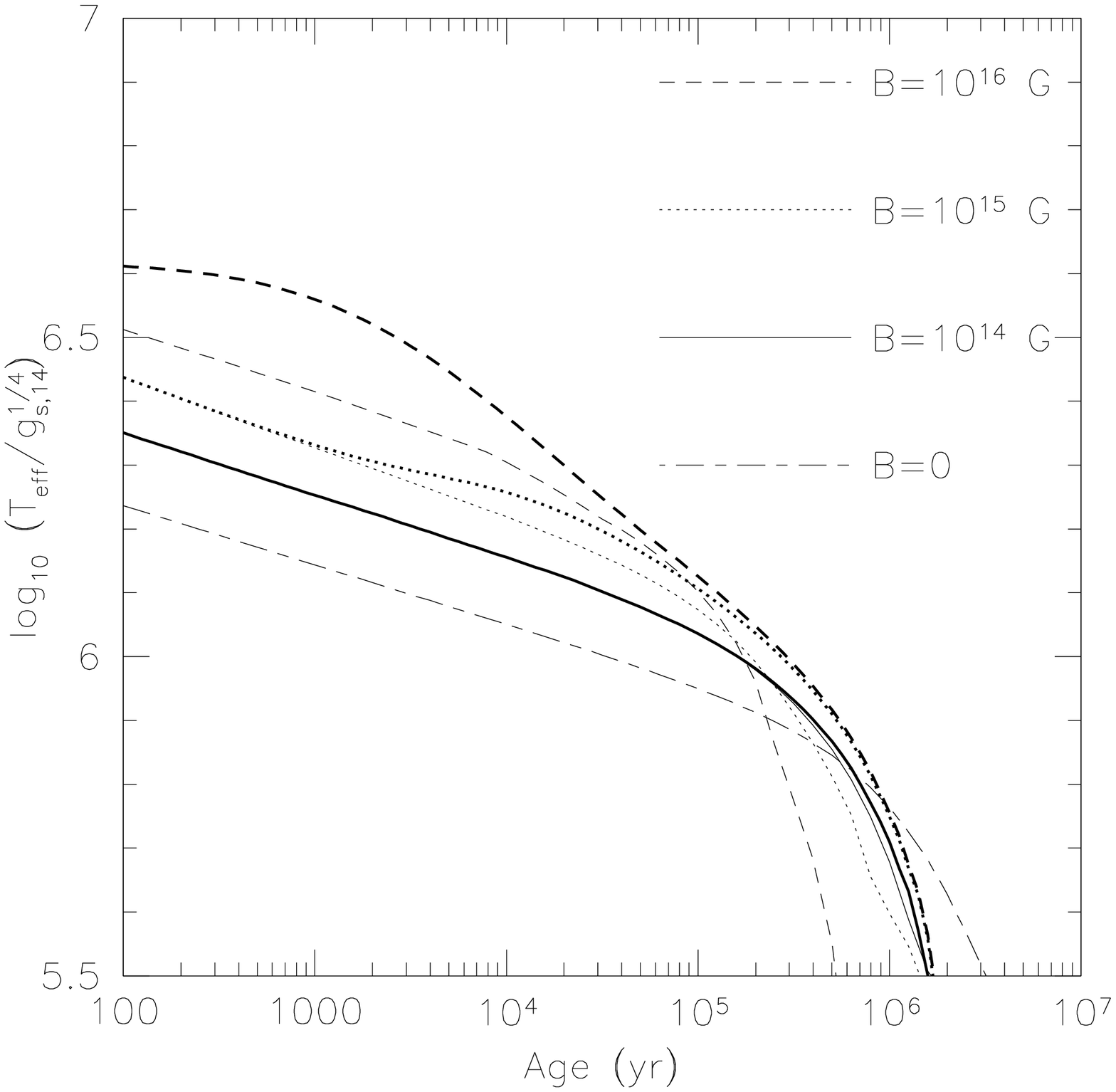}{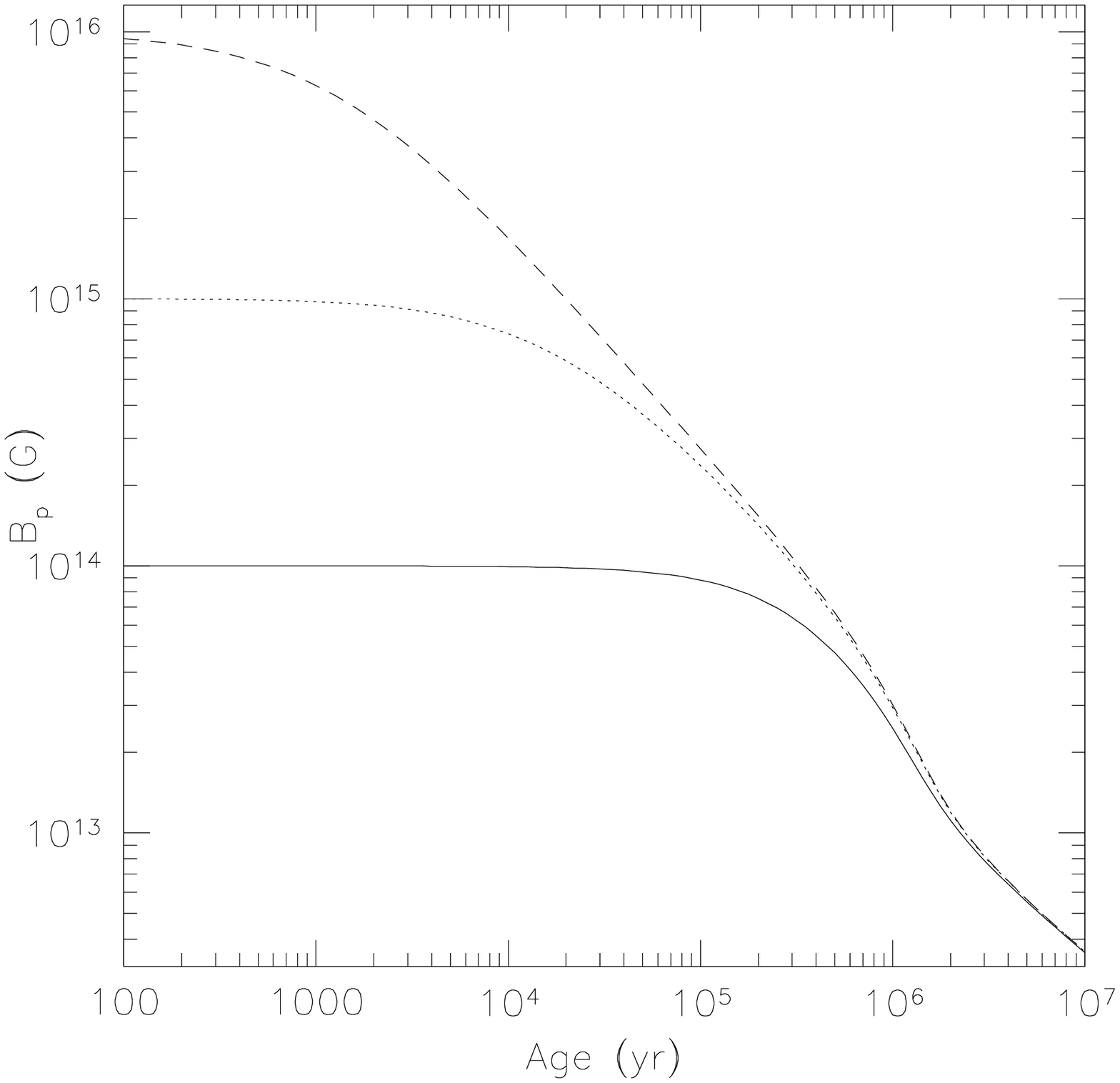}
\plottwo{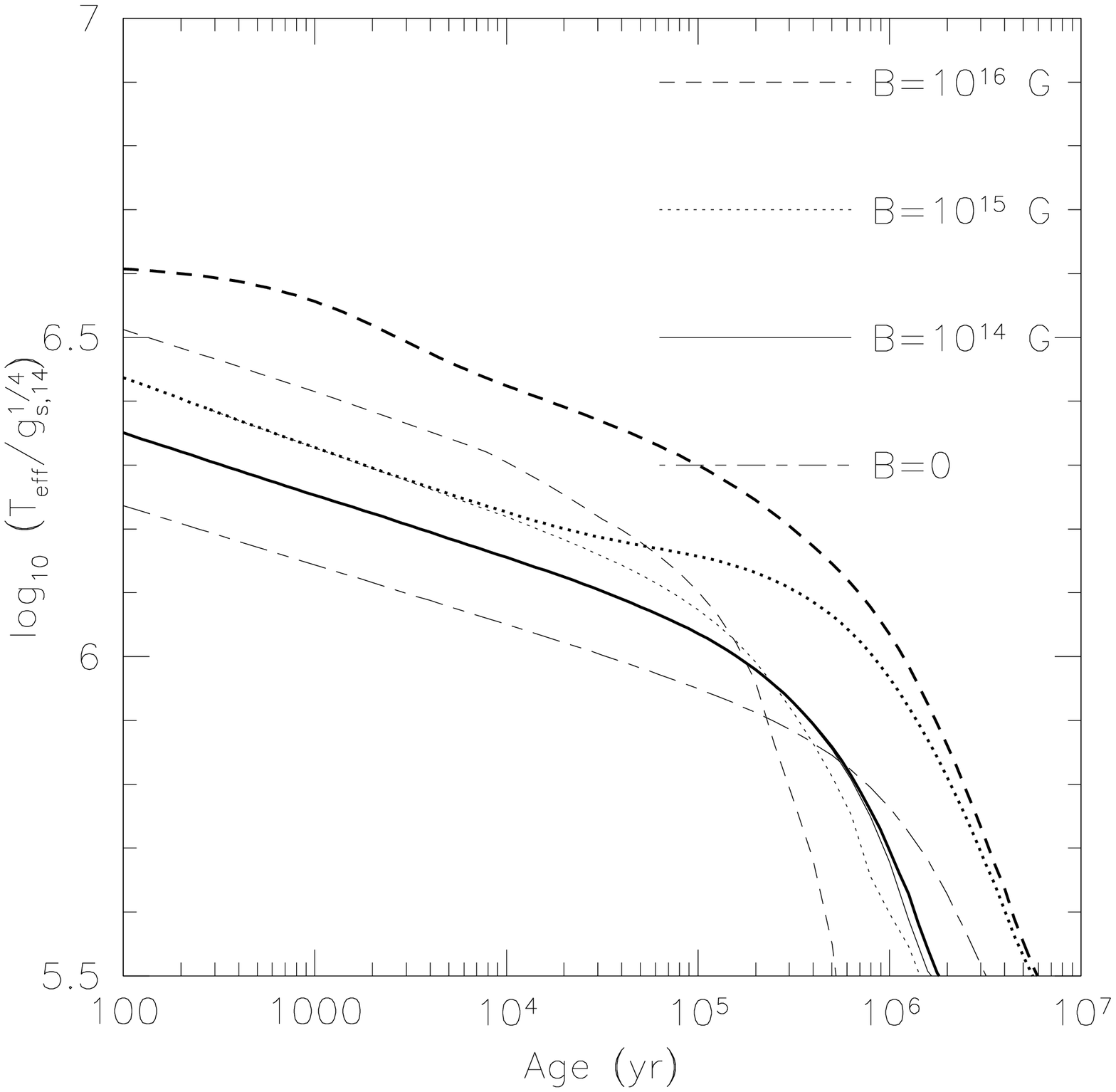}{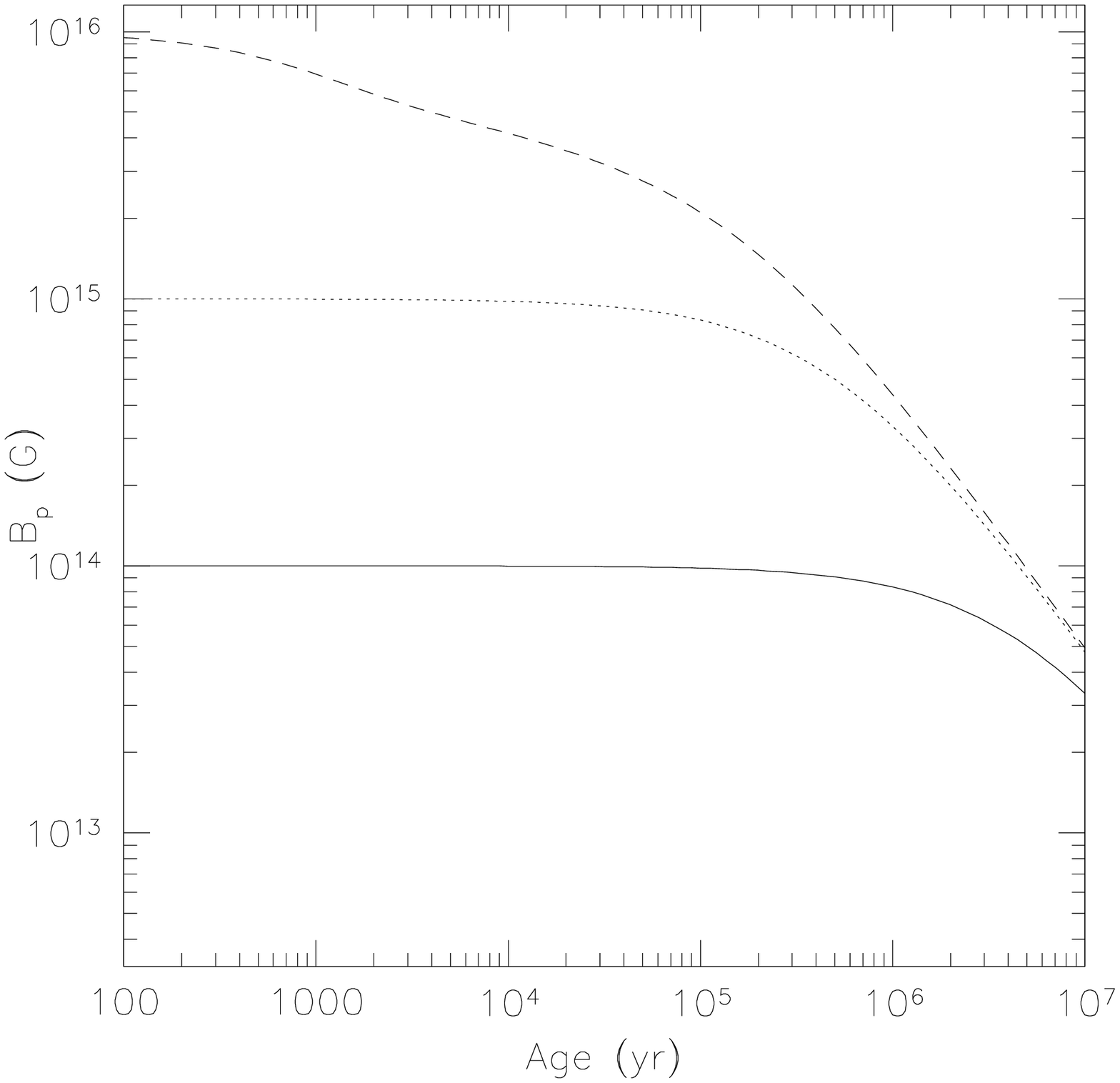}
\caption{The upper panels depict the evolution of the effective
temperature (left) and magnetic field (right) for the case where the
magnetic field decays through the solenoidal mode.
The lower panels depict the results for decay through the irrotational
mode.
The bold solid, dotted and dashed lines give the results for
$B_i=10^{14}, 10^{15}$ and $10^{16}$~G respectively.  The light lines
give the results without any field decay, and the long-short-dashed
line gives the cooling evolution for an unmagnetized neutron star.}
\label{fig:decay}
\end{figure}
}

Our models for the decay and radiation of the magnetic field energy
yield comparable X-ray luminosities to the calculations of
\jcite{Thom96} who discuss the mechanisms for field decay in magnetars
in further detail but extrapolate the envelope models \jcite{VanR88}
which were derived for more weakly magnetized objects ($B \leq
10^{14}$~G) to $B\sim 4.4 \times 10^{15}$~G.  For field decay
dominated by the solenoidal mode, we generally find that photon
cooling begins dominate earlier than for irrotational decay.  For the
irrotational mode our results compare well with \jcite{Thom96}
conclusions.

\section{Discussion}
\label{sec:discuss}

In this paper we have estimated the cooling radiation from magnetars.
Two factors conspire to make them brighter relative to the standard
pulsars. First, the decay of the magnetic field by ambipolar diffusion
(see section~\ref{sec:decay}) is a new source of heat. Second, thermal
conduction is affected by strong magnetic fields
(\cite{Heyl97magnetar}).  The main uncertainties are the rate at which
the field can decay which depends on how the neutron-electron-proton
plasma returns to equilibrium.  If the solenoidal mode operates,
neutron stars with decaying fields of $10^{15}$ to $10^{16}$~G may be
50\% to 250\% more luminous than objects with static fields similar in
magnitude during the neutrino-cooling epoch.  If the irrotational mode
dominates the cooling, the effect is strongest later in the neutron
stars evolution, during the photon-cooling epoch.

Here, we have considered only envelopes consisting of iron.  If
neutron star envelope consists largely of lighter elements (hydrogen
or helium), the opacity of the envelope is even smaller, making the
putative magnetars even brighter (\cite{Heyl97kes}, \cite{Pote97}).
By increasing the transparency of the envelope (\eg \cite{Heyl97kes}),
even a static field of $B\sim 10^{15}$ (weaker than required by
\jcite{Thom96}) can account for the quiescent emission of SGRs.
Regardless of composition though, neutron stars with decaying magnetic
fields will be more luminous than their non-decaying coevals.

We take \rxj\ as a more modest exemplar (as compared to the SGRs) for
magnetar class; it is a remarkably bright source in the ROSAT All-Sky
Survey with a PSPC count rate of 1.67 s$^{-1}$. The distance to this
object is uncertain but it is estimated to lie between 100 to 400 pc.
If we take a nominal limiting flux for the ROSAT All-Sky Survey Bright
Source Catalogue of 0.10 s$^{-1}$ where it has a sky coverage of 92\%,
we would expect $\sim 50$ similar objects to be appear in the
survey. However, the observed population will be restricted severely
by the opacity (photoelectric absorption) of the neutral interstellar
medium. Nonetheless, it is not unreasonable to expect a dozen or so of
objects similar to \rxj\ but fainter.

For the reasons discussed above, in a flux-limited and a
distance-limited sample of soft X-ray sources, middle-aged ($\sim
10^5$~yr) magnetars will be over-represented relative to middle-aged
neutron stars with typical field strengths of $10^{12}$ G (or weaker).
In principle, it is possible to identify this population.  However, in
practice, it is difficult to identify weaker cousins of \rxj\ using
the existing data. Fortunately, AXAF observations of the ROSAT sample
of soft X-ray sources will have the necessary astrometric precision
and spectroscopic resolution differentiate cooling neutron stars from
dominant contaminating populations of white dwarfs, soft AGN and CVs.

Magnetic field decay along with enhanced thermal conduction makes
ultramagnetized neutron stars much brighter than their weakly
magnetized coevals during the neutrino-dominated cooling era;
consequently, the search volume for young neutron stars associated
with supernova remnants is much larger for magnetars with decaying
fields than for weakly magnetized neutron stars.  These unusual
objects may be commonplace in our flux-limited view of the galaxy, and
these processes may explain the quiescent emission of anomalous X-ray
pulsars and soft-gamma repeaters among other phenomena.

\acknowledgements

The authors would like to thank P. Goldreich for useful discussions.
J.S.H. acknowledges a Lee A. DuBridge postdoctoral scholarship.


\begin{thebibliography}{}

\bibitem[\protect{Goldreich \& Reisenegger~\protect\oyear
  1992\protect\cyear}]{Gold92}
Goldreich, P. \& Reisenegger, A. 1992,
\newblock {\em ApJ,} {\bf 395}, 250.

\bibitem[\protect{Haberl et~al.~\protect\oyear 1997\protect\cyear}]{Habe97}
Haberl, F., Motch, C., Buckley, D. A.~H., Zickgraf, F.-J. \& Pietsch, W. 1997,
\newblock {\em A\&A,} {\bf 326}, 662.

\bibitem[\protect{Hernquist~\protect\oyear 1984\protect\cyear}]{Hern84a}
Hernquist, L. 1984,
\newblock {\em ApJS,} {\bf 56}, 325.

\bibitem[\protect{Heyl \& Hernquist~\protect\oyear
  1997\protect\cyeara}]{Heyl97kes}
Heyl, J.~S. \& Hernquist, L. 1997a,
\newblock {\em ApJL,} {\bf 489}, 67.

\bibitem[\protect{Heyl \& Hernquist~\protect\oyear
  1997\protect\cyearb}]{Heyl97magnetar}
Heyl, J.~S. \& Hernquist, L. 1997b,
\newblock {\em ApJL,} {\bf 491}, 95.

\bibitem[\protect{Heyl \& Hernquist~\protect\oyear
  1998\protect\cyeara}]{Heyl98numens}
Heyl, J.~S. \& Hernquist, L. 1998a,
\newblock {\em MNRAS,},
\newblock submitted

\bibitem[\protect{Heyl \& Hernquist~\protect\oyear
  1998\protect\cyearb}]{Heyl98rxj}
Heyl, J.~S. \& Hernquist, L. 1998b,
\newblock {\em MNRAS,},
\newblock in press

\bibitem[\protect{Kouveliotou et~al.~\protect\oyear
  1994\protect\cyear}]{Kouv94}
Kouveliotou, C. {\it et~al.} 1994,
\newblock {\em Nature,} {\bf 368}, 125.

\bibitem[\protect{Kulkarni \& Frail~\protect\oyear 1993\protect\cyear}]{Kulk93}
Kulkarni, S.~R. \& Frail, D.~A. 1993,
\newblock {\em Nature,} {\bf 365}, 33.

\bibitem[\protect{Kulkarni \& {van Kerkwijk}~\protect\oyear
  1998\protect\cyear}]{Kulk98}
Kulkarni, S.~R. \& {van Kerkwijk}, M.~H. 1998,
\newblock {\em ApJ,},
\newblock in press

\bibitem[\protect{Leinson \& Perez~\protect\oyear 1997\protect\cyear}]{Lein97}
Leinson, L.~B. \& Perez, A. 1997,
\newblock {\em Direct URCA process in neutron stars with strong magnetic
  fields},
\newblock astro-ph/9711216

\bibitem[\protect{Motch \& Haberl~\protect\oyear 1998\protect\cyear}]{Motc98}
Motch, C. \& Haberl, F. 1998,
\newblock {\em A\&A,},
\newblock in press

\bibitem[\protect{Narayan~\protect\oyear 1987\protect\cyear}]{Nara87}
Narayan, R. 1987,
\newblock {\em ApJ,} {\bf 319}, 162.

\bibitem[\protect{Potekhin, Chabrier \& Yakovlev~\protect\oyear
  1997\protect\cyear}]{Pote97}
Potekhin, A.~Y., Chabrier, G. \& Yakovlev, D.~G. 1997,
\newblock {\em A\&A,} {\bf 323}, 415.

\bibitem[\protect{Rothschild, Kulkarni \& Lingenfelter~\protect\oyear
  1994\protect\cyear}]{Roth94}
Rothschild, R.~E., Kulkarni, S.~R. \& Lingenfelter, R.~E. 1994,
\newblock {\em Nature,} {\bf 368}, 432.

\bibitem[\protect{Thompson \& Duncan~\protect\oyear
  1995\protect\cyear}]{Thom95}
Thompson, C. \& Duncan, R.~C. 1995,
\newblock {\em MNRAS,} {\bf 275}, 255.

\bibitem[\protect{Thompson \& Duncan~\protect\oyear
  1996\protect\cyear}]{Thom96}
Thompson, C. \& Duncan, R.~C. 1996,
\newblock {\em ApJ,} {\bf 473}, 322.

\bibitem[\protect{Usov \& Melrose~\protect\oyear 1996\protect\cyear}]{Usov96}
Usov, V.~V. \& Melrose, D.~B. 1996,
\newblock {\em ApJ,} {\bf 464}, 306.

\bibitem[\protect{{Van Riper}~\protect\oyear 1988\protect\cyear}]{VanR88}
{Van Riper}, K.~A. 1988,
\newblock {\em ApJ,} {\bf 329}, 339.

\bibitem[\protect{Vasisht \& Gotthelf~\protect\oyear
  1997\protect\cyear}]{Vasi97b}
Vasisht, G. \& Gotthelf, E.~V. 1997,
\newblock {\em ApJL,} {\bf 486}, 129.

\end{thebibliography}

\figcomment{
\end{document}
\end
}
\clearpage

\begin{figure}
\figcaption[ upper left: corete.eps,    upper right coreb.eps;
	    lower left: coreirrte.eps,    lower right coreirrb.eps]
{The upper panels depict the evolution of the effective
temperature (left) and magnetic field (right) for the case where the
magnetic field decays through the solenoidal mode.
The lower panels depict the results for decay through the irrotational
mode.
The bold solid, dotted and dashed lines give the results for
$B_i=10^{14}, 10^{15}$ and $10^{16}$~G respectively.  The light lines
give the results without any field decay, and the long-short-dashed
line gives the cooling evolution for an unmagnetized neutron star.}
\label{fig:decay}
\end{figure}

\end{document}